\documentstyle[12pt,fleqn,twoside,espcrc1,psfig]{article}

\newcommand{\gtsimeq}{\raisebox{-0.6ex}{$\,\stackrel{\raisebox{-.2ex}%
{$\textstyle >$}}{\sim}\,$}}

\begin{document}

\title{\bf Charging Processes for Dust Particles in Saturn's Magnetosphere}

\author{A. L. Graps and E. Gr\"un \address{Max-Planck-Institut f\"ur Kernphysik, Saupfercheckweg 1,
69117 Heidelberg, Germany.}}
\maketitle

\vspace{.5cm}
\begin{center}
\hrule
\end{center}
\vspace{.5cm}

\begin{abstract}
We investigated the electrostatic charging behavior of
submillimeter-sized dust particles located in Saturn's magnetosphere.
The charging effects we considered included electron/ion capture from
the magnetospheric plasma, electron/ion capture from the solar-wind
plasma, the photoelectric effect from solar radiation, and secondary
electron emission from energetic electrons. In our results, we show
charging times and equilibrium potentials for particles located in
different regions of Saturn's magnetosphere. We find that charging in
Saturn's magnetosphere is not particularly sensitive to the dust
particle's material properties.  The equipotential ranges from $\sim$-2~V at
3.5~R$_S$, decreasing to $\sim$-5~V at 6~R$_S$, and then increasing to
$\sim$-1.5~V at 10~R$_S$. The charging time for one micron-sized
particles is a few minutes, and for 0.01 micron-sized particles the
charging time is 6 hours (or more). The latter is a significant
fraction of Saturn's rotation period.
\end{abstract}

\vspace{.5cm}
\begin{center}
\hrule
\end{center}
\vspace{.5cm}

The dynamics of dust, submillimeter-sized dust grains, is a fascinating area of
study of our solar system's dynamical evolution. Small particles, especially
charged particles, respond to other forces than gravitational, in particular,
electromagnetic forces. The dust particle's properties and dynamics fall
into a complex regime between nuclear physics and electromagnetic physics and
gravitational physics.  In order to calculate charges on a dust
particle around a planet, we must characterize:
\newpage

\begin{itemize}
\item The planet's magnetospheric features: its magnetic field and plasma,
\item The physical processes onto the dust particle that generate currents, and
\item The material properties of the dust particle.
\end{itemize}

In this paper, we make reference to results of charging
of dust particles in Earth's magnetosphere (\cite{Juh97},\cite{Graps2000}).
Earth is an interesting charging environment for dust particles, in part, because
the dynamic magnetospheric plasma shows steep changes in the
electron and ion energies and densities, therefore, the electron energy can
be quite high (e.g. a few thousand eV). Table 1 lists some
basic parameters comparing Saturn's and Earth's magnetosphere.

\begin{table}[htp]
\caption {Basic Parameters of Saturn's and Earth's Magnetosphere}
\begin{tabular}{lcc}\hline
Parameters & Saturn & Earth \\ \hline
Rotation Period (day)&  0.44 & 1.0 \\
Dipole Moment (Gauss-cm$^3$) & 2.4 10$^{28}$ & 7.9 10$^{28}$ \\
Field at Equator (Gauss) & 0.22 & 0.305 \\
Dipole Axis (deg) & 0.0 & +10.8 \\
Magnetopause Distance & 20 R$_S$ & 10 R$_E$ \\
Plasma Source & Solar Wind, Atmosphere, & Solar Wind, Atmosphere \\
&  Rings, Moons & \\ \hline
\end{tabular}
\\ Data from \cite{Beatty99}.
\end{table}

\section{Saturn's Plasma Environment}

The electrostatic potential of a dust particle not only depends on
the physical properties of the particle, but also on the plasma
environment, such as the plasma number density, temperature (energy), velocity
distribution of the plasma particles, intergrain distance, and the relative motion
between the dust particles and the plasma (\cite{Kimura98}).

Inside of Saturn's plasmasphere, the plasma density increases towards the planet from
 $\sim$~1 electron per cm$^{-3}$ at Saturn radius 10~R$_S$ to $\sim$~100 electrons
per cm$^{-3}$ at 3~R$_S$, and the electron energy $kT_e$ decreases from
 $\sim$~100~eV to  $\sim$~10~eV.

To characterize Saturn's plasma, we utilized plasma data from
M. Hor\'anyi.  This plasma data is a four component plasma (hydrogen,
oxygen, hot electrons, and cold electrons) fit to the Voyager data
described in \cite{Richardson90}.  The Debye screening length is the
distance that the Coulomb field of an arbitrary charge of the plasma
is shielded. We can calculate the charge for an isolated grain if we
have only one grain within a sphere of radius Debye length.  Figures 1 and
2 show the energy, density our plasma data, and we state, in Table 2,
some representative plasma values for our plasma data.

\begin{table}
\caption {Saturn Plasma Representative Numbers}
\begin{tabular}{lcccc}\hline
Component & Location (R$_S$) & Energy (eV)  & Density (cm$^{-3}$) & Debye Length (m) \\ \hline
Cold electrons & 10 & 8.6 & 1 & 22 \\
&  3 & 0.005 & 52 & 0.073 \\
Hot electrons & 10 & 862 & 0.6 &  280 \\
&  3 & 27 & 0.6 & 50 \\
Hydrogen ions & 10 & 17 & 0.3 & 56 \\
&  3 & 6 & 6 & 7.4 \\
Oxygen ions & 10 & 250 & 0.9 &  120 \\
&  3 & 31 & 46 & 6.1 \\ \hline
\end{tabular}
\end{table}

%Figure 1 goes here
%\newpage
%\thispagestyle{empty}
%\psnodraftbox
%\psfig{figure=/home/graps/iaudust/paper/fig1.ps}

%Figure 2 goes here
%\newpage
%\thispagestyle{empty}
%\psnodraftbox
%\psfig{figure=/home/graps/iaudust/paper/fig2.ps}

\section{Charging Processes}

We calculate the time-varying charge due to currents acting on a
dust particle in a planetary magnetosphere using the following expression:
\begin{equation}
\sum\limits_{k}I_k = I_{\rm i,e,moving} + I_{\rm sec} + I_\nu
\end{equation}
where $I_k$ is the current of the $k$-th charging processes
(\cite{Kimura98}).  We consider 3 charging currents. The first
charging current is: $I_{\rm i,e,moving}$, which is the collection of
ions and electrons onto the dust particle from the ambient plasma. The
second current: $I_{\rm sec}$, the secondary electron current, occurs
when a high energy electron impacts the dust particle, some of the
dust material is ionized, and electrons are ejected from the
particle. The third current, $I_\nu$, photoelectron emission current,
occurs when a UV photon impacts the dust particle and photoelectrons
are released.  For a more complete treatment, one should add reflected
electrons from the secondary electron emission (\cite{Jurac95}) and
the small particle effect (\cite{Draine79}).

The secondary electron current is dependent on the dust particle
material. If one wants to characterize different dust material
properties, then one applies the secondary electron emission maximum
yield $\delta_{\rm m}$ and the primary energy $E_{\rm m}$ at which the
maximum yield occurs, acquired from laboratory measurements. The yield
is the ratio of the secondary current to the primary current, given
the energy of the impacting electron or ion. Example yield and energy
values for relevant solar system material is shown in Table 3.

\begin{table}
\caption {Examples of Dust Particle Material Properties}
\begin{tabular}{lccc}\hline
Material & density (g-cm$^3$) & $\delta_{\rm m}$ &  $E_{\rm m}$ (eV) \\ \hline
Graphite & 2.26 & 1 & 250 \\
SiO$_2$ & 2.65 & 2.9 & 420 \\
Mica & 2.8 & 2.4 & 340 \\
Fe & 7.86 & 1.3 & 400 \\
Al & 2.70 & 0.95 & 300 \\
MgO & 3.58 &23 & 1200 \\
Lunar dust & 3.2 & $\sim$1.5 & 500 \\ \hline
\end{tabular}
\\ Data from \cite{Draine79}.
\end{table}

\section{Charging Results}

We choose, as our canonical example, a hybrid dust particle with
material properties similar to a conducting graphite particle
$\delta_{\rm m}$=1.5, $E_{\rm m}$ = 250 eV, but with photoelectron
yield properties similar to a dielectric particle (in Hor\'anyi et
al's, modeling work, the photoelectron yield is denoted $\chi$ and
ranges from $\chi$=1.0 for conducting magnetite dust particles to
$\chi$=0.1 for dielectric olivine particles). These properties were
chosen in order to compare with charging results we have obtained for
a dust particle in Earth's magnetosphere.

For our canonical dust particle, we calculated the equilibrium
potential (``equipotential''), the charging time, and examined the dominant currents for a
particle at Saturn radii locations of 3~R$_S$ to 10~R$_S$. Equilibrium
potential for the dust particle is reached when the sum of the
charging currents is zero. The charging time is the time for a
particle's potential to reach an equilibrium. The currents that we
examined are the electron and ion collection currents, the
photoelectron current and the secondary electron current. Figures
3a,b,c display the results for the equilibrium potential, the charging
time, and the dominant currents for our canonical case, 1~$\mu$m dust
particle.  Here, the equipotential ranges from $\sim$-2~V at
3.5~R$_S$, decreasing to $\sim$-5~V at 6~R$_S$, and then increasing to
$\sim$-1.5~V at 10~R$_S$.  The charging time for the starred positions
is $\sim$~1~minute. If we perform the same calculations for a 100
times smaller particle with the same material properties, then we find
charging times on order of a few hours, which is a signficant fraction
of Saturn's rotation period. Also, for a 100 times smaller particle, the
secondary electron emission current will be more efficient, causing
the smaller particle to charge more positively than for the larger (1 ~$\mu$m-sized) dust
particle.

%Figure 3
%\newpage
%\thispagestyle{empty}
%\psnodraftbox
%\psfig{figure=/home/graps/iaudust/paper/fig3.ps}

For an identical dust particle in a geostationary location in Earth's magnetosphere,
we calculated dramatic differences in the equipotential values and the charging
times when we applied plasma conditions appropriate to ``disturbed" and
``quiet" Earth magnetosphere conditions, and when we slightly varied
the material properties from
$\delta_{\rm m}$=1.5, E$_{\rm m}$ = 250 eV, $\chi$=0.1 to
$\delta_{\rm m}$=1.4, E$_{\rm m}$ = 180 eV, $\chi$=0.1. For the first
set of material properties in quiet Earth plasma conditions, the equilibrium potential was
$\sim$+5~V. However, for the second set of material properties in
disturbed Earth plasma conditions, the equilibriuim potential was
$\sim$-3000~V. The charging time was about 10 seconds for a
1~$\mu$m dust particle in quiet Earth plasma conditions, and
one-third that time for a 1~$\mu$m dust particle in active Earth plasma
conditions. The charging time generally {\it increases} with
{\it decreasing} particle radii.

What happens when we vary the material properties $\delta_{\rm m}$,
E$_{\rm m}$, and $\chi$=0.1 for a 1~$\mu$m dust particle in the same
way in Saturn's magnetosphere, using the same charging processes as we
applied for a particle in Earth orbit?  Surprisingly, we find very
little change in the resulting potentials, charging times, and
currents, as seen in Figures 4 and 5. Removing each of the currents,
one by one, however, {\it does} have an effect, in particular the
secondary electron emission. If we calculate equipotentials without
the secondary electron emission current, then the dust particle
potential stays negative throughout the magnetosphere, and doesn't
reach positive potentials beyond 8~R$_S$.

On the other hand, removing the photoelectron emission current doesn't
alter the equilibrium potentials in a significant way.

The charging time for a 1~$\mu$m dust particle is on the order of a
few minutes, while for a 0.01~$\mu$m dust particle, the charging time
is on the order of {\gtsimeq few hours}.

%Figure 4 goes here
%\psnodraftbox
%\psfig{figure=/home/graps/iaudust/paper/fig4.ps}

%Figure 5 goes here
%\psnodraftbox
%\psfig{figure=/home/graps/iaudust/paper/fig5.ps}

\section{Summary}

\begin{itemize}
\item Charging in Saturn's magnetosphere is not particularly sensitive to the dust particle's
material properties. This is a large contrast to dust particles in
Earth's magnetosphere, where small material property changes can have a big effect on the
equilibrium potential.
\item The charging time for one micron-sized particles is a few minutes, and for 0.01 micron-sized
particles the charging time is 6 hours (or more). The latter is a signficant fraction
of Saturn's rotation period.
\end{itemize}

\vspace{0.25cm}
{\bf\noindent Acknowledgements.} We are very grateful to
M. Hor\'anyi,  Laboratory for Atmospheric and Space Physics,
University of Colorado, Boulder, for his Saturn plasma data and
helpful guidance on dust particle charging mechanisms.

%Figure Captions
\newpage
\vspace{0.25in} Figure 1: Saturn plasma energies of our four component plasma:
hot electrons, cold electrons, oxygen ions, and hydrogen ions.

\vspace{0.25in} Figure 2: Saturn plasma densities of our four component plasma:
hot electrons, cold electrons, oxygen ions, and hydrogen ions. The highest energy
and density components are the oxygen ions and hot electrons.

\vspace{0.25in} Figure 3: a) Equilibrium potential (V), b) Charging time (sec),
and c) Currents (e s$^{-1}$cm$^{-2}$) for a 1~$\mu$m dust particle of material
properties $\delta_{\rm m}$=1.5, E$_{\rm m}$ = 250 eV, $\chi$=0.1.

\vspace{0.25in} Figure 4: a) Equilibrium potential (V),
and b) Currents (e s$^{-1}$cm$^{-2}$) for a 1~$\mu$m dust particle of material
properties $\delta_{\rm m}$=2.4, E$_{\rm m}$ = 400 eV, $\chi$=0.1.

\vspace{0.25in} Figure 5: a) Equilibrium potential (V),
and b) Currents (e s$^{-1}$cm$^{-2}$) for a 1~$\mu$m dust particle of material
properties $\delta_{\rm m}$=1.4, E$_{\rm m}$ = 180 eV, $\chi$=0.1.

%Figure 1
\newpage
\thispagestyle{empty}
\psnodraftbox
\psfig{figure=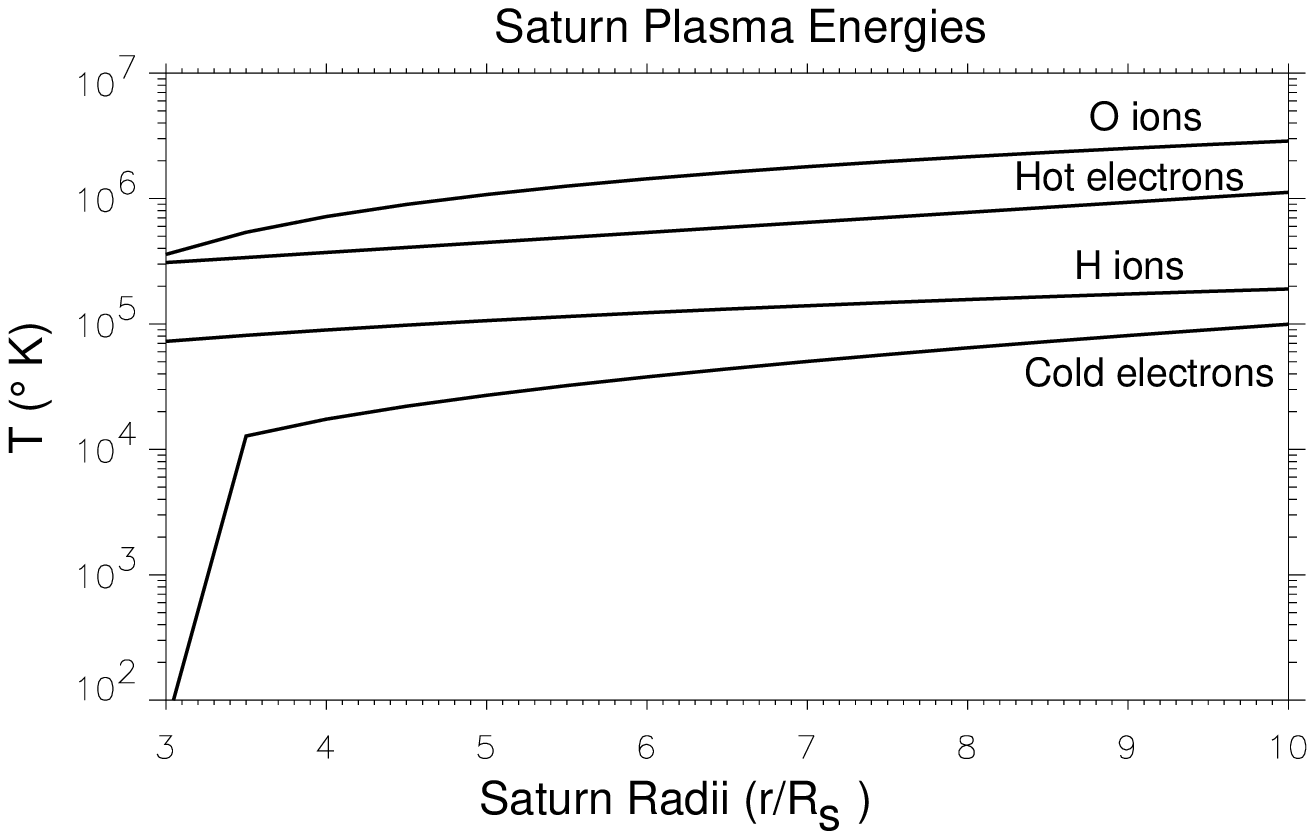}

%Figure 2
\newpage
\thispagestyle{empty}
\psnodraftbox
\psfig{figure=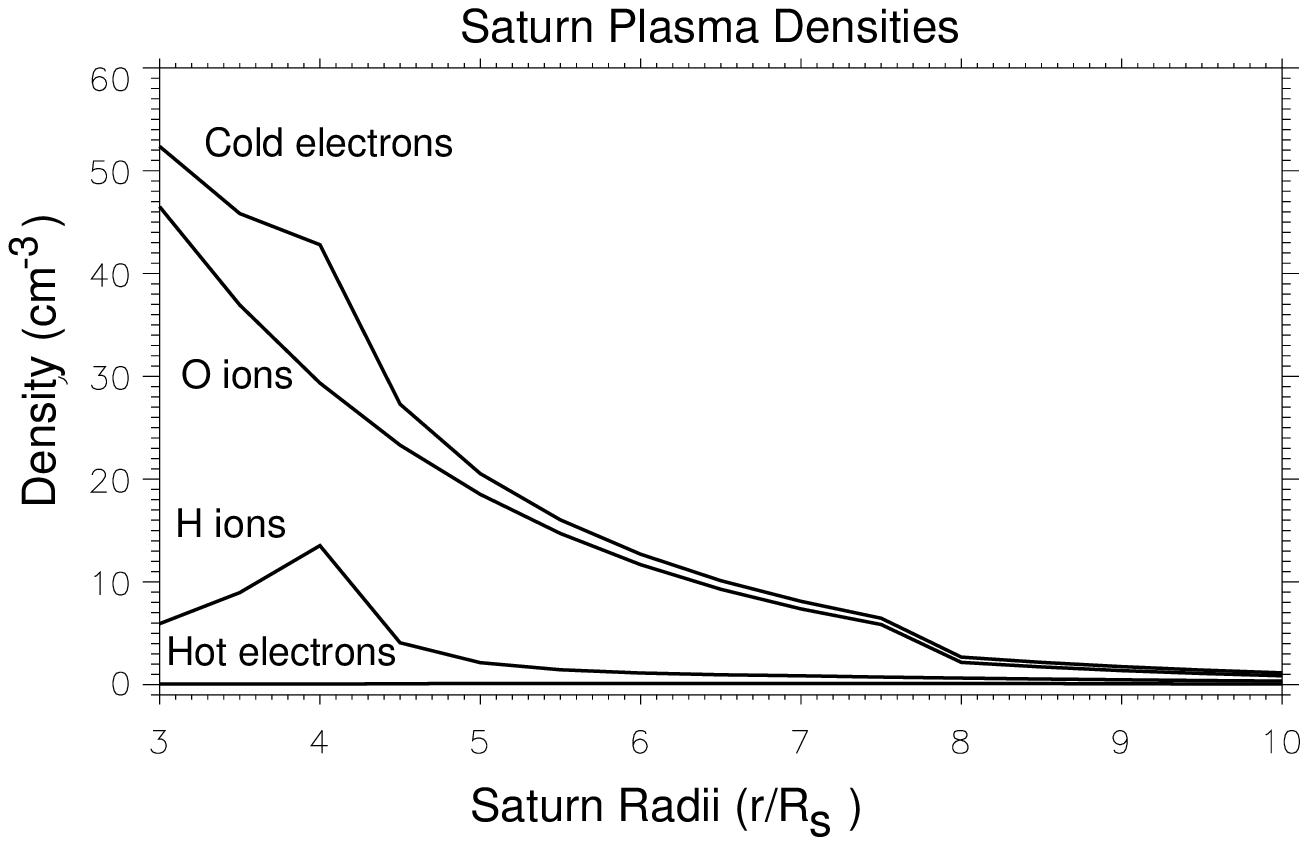}

%Figure 3
\newpage
\thispagestyle{empty}
\psnodraftbox
\psfig{figure=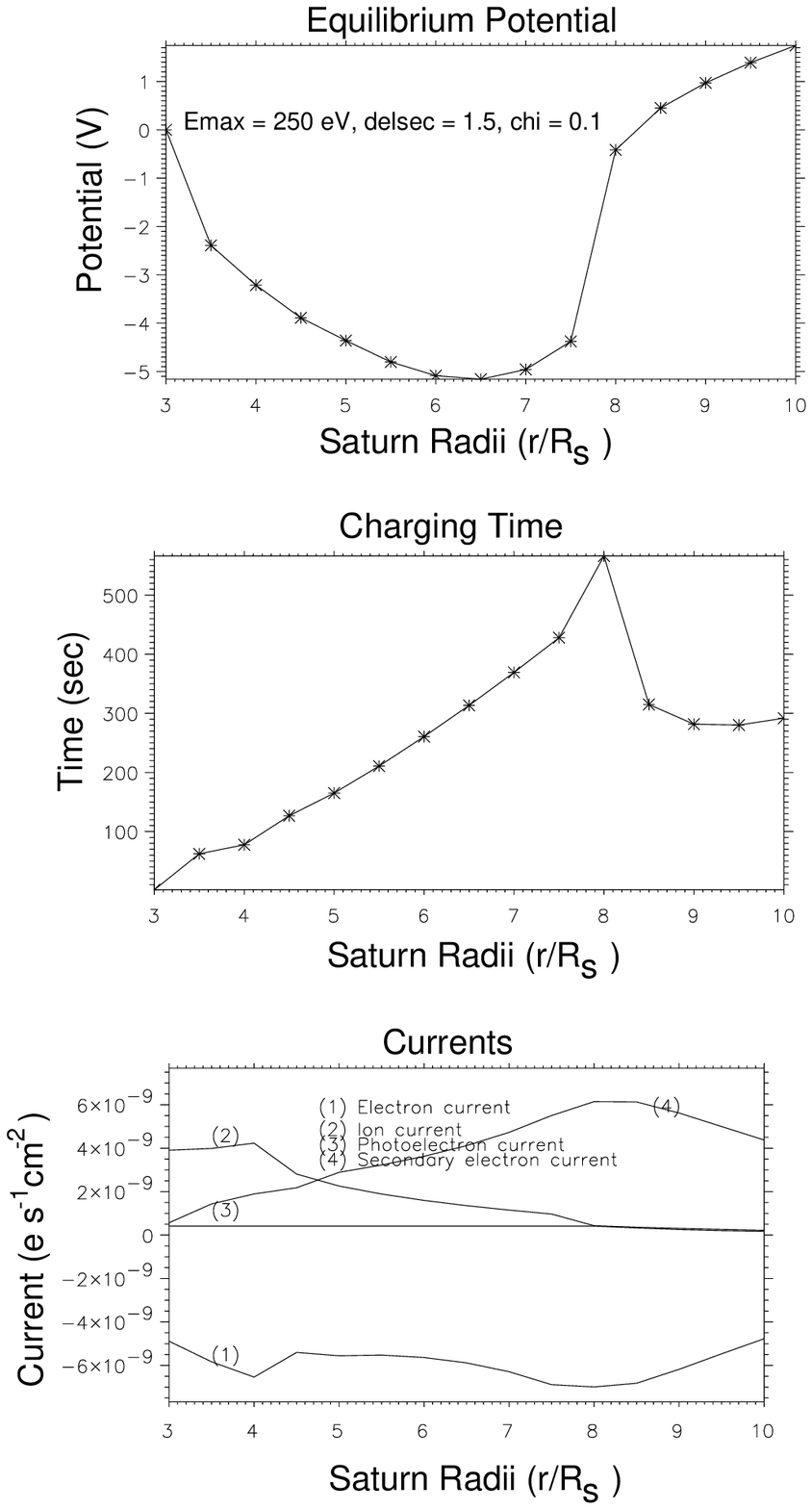}

%Figure 4
\newpage
\thispagestyle{empty}
\psnodraftbox
\psfig{figure=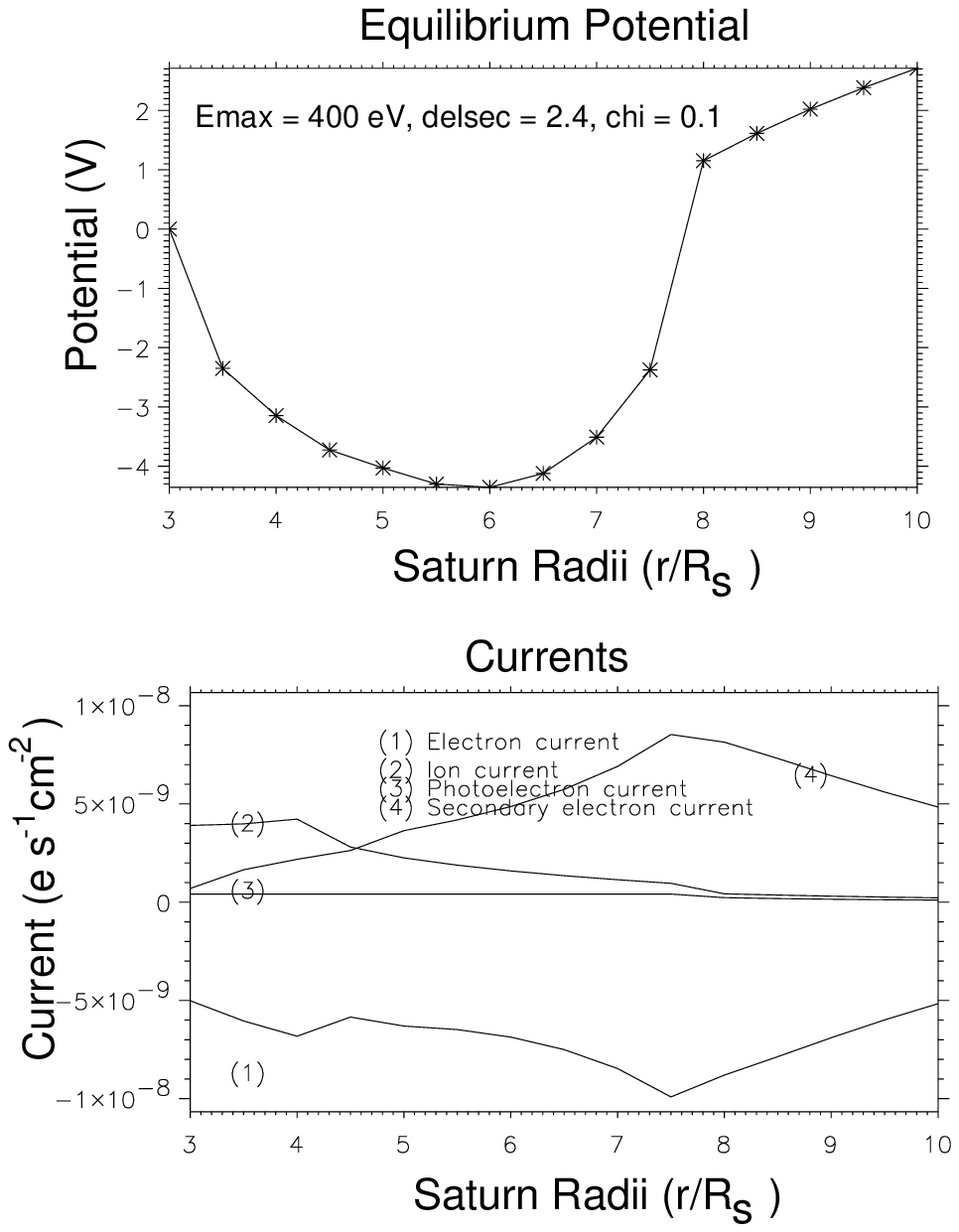}

%Figure 5
\newpage
\thispagestyle{empty}
\psnodraftbox
\psfig{figure=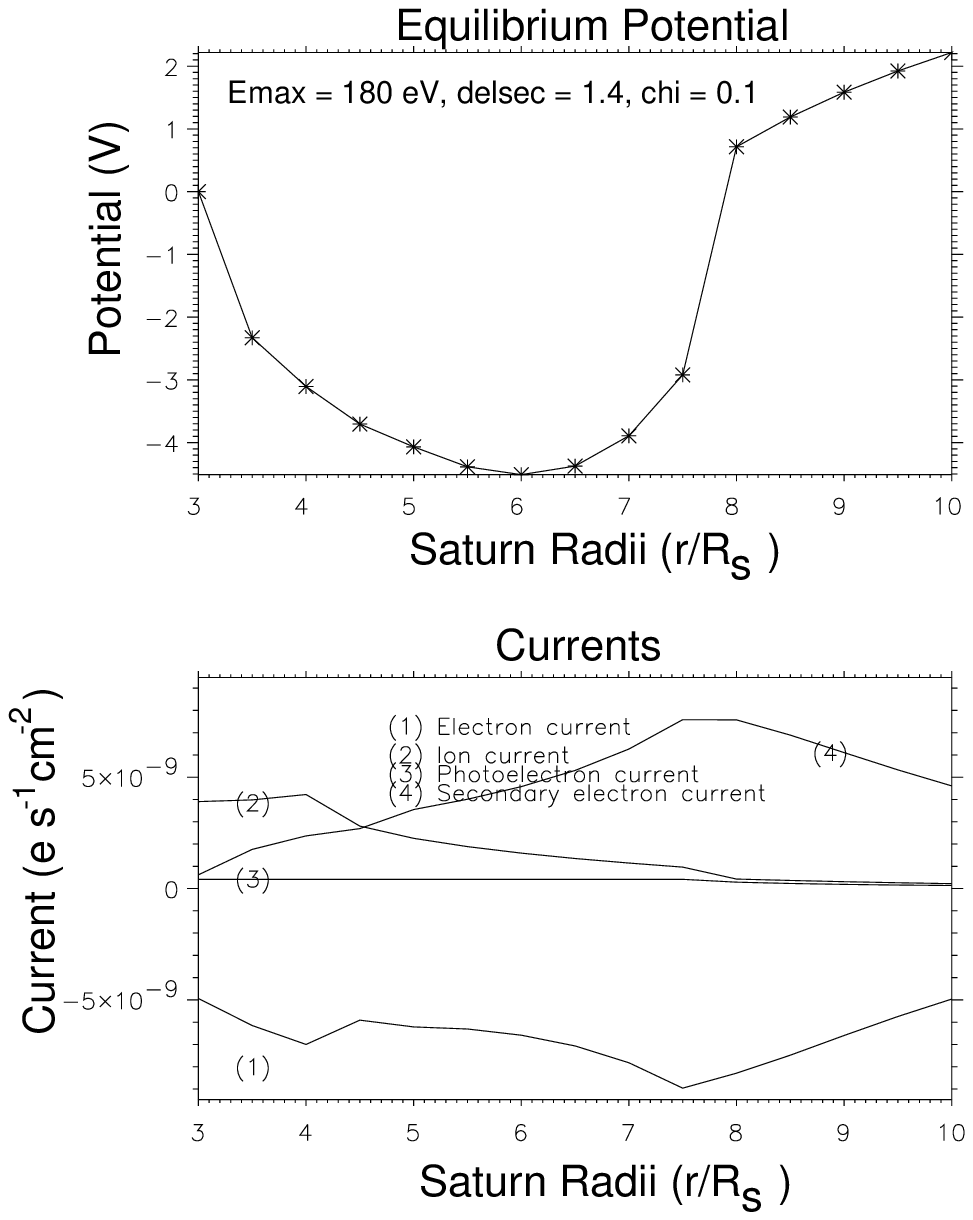}

% End of document.
\end{document}